\documentstyle[prd,aps,floats]{revtex}

\newcommand{\be}{\begin{equation}}
\newcommand{\ee}{\end{equation}}
\newcommand{\bea}{\begin{eqnarray}}
\newcommand{\eea}{\end{eqnarray}}

\begin{document}

\draft

\renewcommand{\topfraction}{0.8}
\renewcommand{\bottomfraction}{0.8}
\twocolumn[\hsize\textwidth\columnwidth\hsize\csname
@twocolumnfalse\endcsname

\title{Supergravity Brane Cosmologies}

\author{James E. Lidsey}

\address{Astronomy Unit, School of Mathematical Sciences, Queen 
Mary and Westfield,
Mile End Road, London, E1 4NS, UK}

\maketitle
\begin{abstract}

Solitonic brane cosmologies are found 
where the world-volume is curved due to 
the evolution of the dilaton field on the brane. 
In many cases, these 
may be related to the solitonic D$p$- and M5-branes of string and M-theory.
An eleven-dimensional interpretation 
of the D8-brane cosmology of the massive type IIA 
theory is discussed in terms 
of compactification on a torus bundle.
Braneworlds  are also found in 
Horava-Witten theory compactified on a Calabi-Yau three-fold.  
The possibility of dilaton-driven inflation on the brane is discussed.

\end{abstract}


\vskip2pc]

\section{Introduction}

\setcounter{equation}{0}

There has been considerable interest recently in
the possibility that our observable  
universe may be viewed as a $p$--brane 
embedded in a higher--dimensional 
spacetime. In this picture, the gauge interactions 
are confined to 
the brane, but gravity may propagate in the bulk. 
This change in viewpoint has been partially motivated 
by advances in our understanding 
of non--perturbative string theory. 
For example, the strongly--coupled, 
field theoretic limit 
of the ${\rm E}_8 \times {\rm E}_8$ 
heterotic string
has been described by  
Ho\v{r}ava and Witten as $D=11$ supergravity 
on an orbifold ${\rm S}^1/{\rm Z}_2$, where the two 
sets of ${\rm E}_8$ 
gauge fields are confined to the orbifold fixed planes \cite{hw}. 
Compactification of this theory on a Calabi--Yau three--fold 
results in an effective five--dimensional theory containing 
a superpotential \cite{WittenE,LOSW} that 
supports a pair of 
parallel  3--branes (domain walls) \cite{LOSW}. 

A related five--dimensional model with an extra orbifold
dimension was recently proposed 
by Randall and Sundrum. In this model, 
there are two 3--branes with equal 
and opposite tensions at the orbifold fixed points and 
our universe is identified as the positive tension 
brane. The existence of a negative  
cosmological constant  in the 
five--dimensional bulk results in a curved  
background. This supports a bound state of 
the higher--dimensional graviton 
that is localized to the 
3--brane \cite{RS} 
and, consequently, the size of the 
extra dimension can  be 
arbitrarily large. 

This picture differs significantly from traditional 
Kaluza--Klein compactification, 
where the higher--dimensional universe is represented 
as a product space and the four--dimensional 
Planck mass is determined by the volume of the 
extra dimensions. In the braneworld scenario, the geometry is 
{\em non--factorizable} because the brane tension 
induces 
a `warp factor' in the metric. 

A crucial question that must be addressed is whether the
braneworld scenario is 
consistent with  our understanding of 
early universe cosmology. A central paradigm 
of the early universe is that of 
cosmological inflation, where the universe undergoes 
an epoch of accelerated expansion. 
It is therefore 
important to develop brane cosmologies where 
inflation may proceed. 

The cosmological implications of the 
Randall--Sundrum model have been considered
by a number of authors \cite{oned}
and other examples of curved 
branes  were recently  presented in Ref. \cite{morecurved}. 
In view of the above 
developments, we show in this paper that a wide class of 
higher--dimensional supergravity theories admit 
solutions that may be interpreted as (non--supersymmetric) 
brane cosmologies, where
the dilaton field varies non--trivially over the 
world--volume. The 
effect of this variation is formally equivalent, 
after appropriate field redefinitions, to the 
introduction of a massless, minimally coupled
scalar field on the brane. 
Hence, these solutions are 
relevant to inflationary 
models based on string 
theory such as the pre--big bang scenario \cite{pbb}, 
since, in 
this model, the accelerated expansion is driven by 
the kinetic energy of the dilaton field.

The paper is organised as follows. 
The class of brane cosmologies is derived in Section  
II. 
In Secion III, we discuss 
some of the models from an eleven--dimensional 
perspective. Brane cosmologies 
in the Ho\v{r}ava--Witten heterotic theory are found 
in Section IV and we conclude with a discussion 
on dilaton--driven inflation in Section V.

\section{Brane Cosmology in Supergravity Theories}

\setcounter{equation}{0}

\subsection{Ricci--Flat Branes}

We consider the class of $D$--dimensional 
effective actions,
where the graviton, $g_{MN}$, is coupled to 
a dilaton field, $\Phi$, and the
$q$--form  field strength, $F_{[q]}=d A_{[q-1]}$, of an antisymmetric 
gauge field, $A_{[q-1]}$: 
\begin{equation}
\label{action}
S=\int d^D x \sqrt{|g|} \left[ R -\frac{1}{2} 
\left( \nabla \Phi \right)^2 -\frac{1}{2q!} 
e^{\alpha \Phi} F^2_{[q]} \right] , 
\end{equation}
where $R$ is the Ricci  curvature scalar of 
the spacetime, $g \equiv {\rm det}g_{AB}$ and 
the coupling parameter, $\alpha$, is a constant. 
Given appropriate conditions on the form fields, 
action (\ref{action}) represents a consistent truncation
of the bosonic sectors of the $D=10$, $N=2$  
supergravity theories\footnote{Throughout this paper, 
the Chern--Simons terms that also arise in the effective actions 
are trivial for the solutions we consider and we do not 
present them here.}.
The toroidal compactification of the type II theories 
to $D< 10$ also results in an 
action of the form given in Eq. (\ref{action}) 
\cite{LP}. The effective 
action for M--theory is given by Eq. (\ref{action}), 
where $D=11$, $q=0$ and $\Phi=0$ \cite{Witten}.

The field equations derived by varying the action (\ref{action}) 
are\footnote{In 
this paper, upper case, Latin indices take values 
in the range $A=(0, 1, \ldots , D-1)$, 
lower case, Greek indices vary from $\mu =(0, 1, \ldots , 
d-1)$ and lower case, Latin indices from $a=(d, \ldots , D-1)$. 
The dimensionality of the world--volume of the brane is denoted $d_W$, 
the dimensionality of the transverse space is $d_T$  and the 
spacetime metric has signature $(-, +, \ldots , +)$.} 
\begin{eqnarray}
\label{field1}
R_{AB} =\frac{1}{2} \nabla_A \Phi \nabla_B \Phi
 + 
\frac{1}{2q!} e^{\alpha \Phi} \left( q F_{AC_2 \ldots C_q} 
{F_B}^{C_2 \ldots C_q} \right. \nonumber \\
\left. -\left( \frac{q-1}{D-2} \right)
F^2_{[q]} g_{AB} \right) \\
\label{field2}
\nabla^2 \Phi = \frac{\alpha}{2q!}e^{\alpha \Phi} F^2_{[q]} \\
\label{field3}
\nabla_A \left( e^{\alpha \Phi} F^{AB_2 \ldots B_q} \right) =0 .
\end{eqnarray}
When $q=0$, the field strength  may be interpreted as a 
cosmological constant. For $q=1$, it 
represents the gradient of a
massless axion field. 

In $D$ dimensions, a solitonic $(D-q-2)$--brane is 
supported by the `magnetic' charge of a  $q$-form field strength. 
Thus, M--theory contains a 5--brane 
due to the four--form field strength (M5--brane).
Moreover, both ten--dimensional type II theories  
admit a 5--brane supported by the 
Neveu--Schwarz/Neveu--Schwarz (NS--NS) three--form 
field strength (NS5--brane)  \cite{DGHR}. 
These theories also admit
branes supported by Ramond-Ramond (RR) fields (D$p$--branes). 
(For a review, see, e.g., Ref. 
\cite{Argurio}). 
The RR sector 
of the type IIB theory contains a one--form, a three--form and 
a (self--dual) five--form that result in a 
D7--, D5-- and D3--brane, respectively. 
The massless type IIA theory, on the other 
hand, admits a D6-- and D4--brane, whereas 
the massive type IIA theory 
due to Romans \cite{Romans} also admits a D8--brane supported by a 
0--form. In general, 
the coupling 
of the RR $q$---forms to the dilaton in the 
type II theories 
is given by $\alpha =(5-q)/2$. 

Recently, 
Brecher and Perry \cite{Brecher} showed that 
Eqs. (\ref{field1})--(\ref{field3}) admit solitonic 
D$p$-- and M--brane solutions with a Ricci--flat 
world--volume, $f_{\mu\nu} =f_{\mu\nu}(x^{\rho})$: 
\begin{eqnarray}
\label{brane1}
ds^2_D =H^m f_{\mu\nu}dx^{\mu}dx^{\nu} +H^n \delta_{ij} 
dy^idy^j \\
\label{brane2}
e^{\Phi} =H^{-\alpha /2} \\
\label{brane3}
F_{a_1 \ldots a_q} = \lambda 
\epsilon_{a_1 \ldots a_q b}\frac{y^b}{r^{q+1}} ,
\end{eqnarray}
where  
\begin{equation}
\label{mn}
m=-\frac{q-1}{D-2}, \quad 
n=\frac{D-q-1}{D-2} 
\end{equation}
and $d_T=q+1$. The coordinates on the space transverse to 
the brane are $\{ y^i \}$ and $H=H(r)$ is an harmonic 
function on this space:
\begin{equation}
\label{harmonic} 
\delta^{ij} \partial_i \partial_j H=0 ,
\end{equation}  
where $r$ represents the radial 
coordinate. The components of the 
alternating tensor in Eq. (\ref{brane3}) are $\pm 1$ and 
$\lambda$ is a constant. 
The solutions preserve some fraction of the 
supersymmetry if the world--volume admits parallel 
spinors \cite{ff}.  For the M5--brane, $m=-1/3$, 
$n=2/3$ and $\alpha =0$.

\subsection{Curved Branes}

In this paper we allow the transverse space to 
depend directly  on the 
world--volume coordinates
of the brane by introducing a 
scalar function $B=B(x^{\mu})$: 
\begin{equation}
\label{branecos}
ds^2_D =H^m f_{\mu\nu} dx^{\mu}dx^{\nu} +e^{2B}H^n \delta_{ij} dy^i
dy^j  ,
\end{equation}
where $(m,n)$ and $H$ are defined 
in Eqs. (\ref{mn}) and (\ref{harmonic}), 
respectively. 
In the standard 
Kaluza--Klein picture, 
the degree of freedom, $B$, would represent
the  `breathing mode' of the 
internal dimensions and  would play
the role of a modulus field 
in the lower--dimensional theory. 

The components of the Ricci tensor for the 
metric (\ref{branecos}) are  given by
\begin{eqnarray}
R_{\mu\nu} = \bar{R}_{\mu\nu} -d_T \left( 
\bar{\nabla}_{\mu\nu}B +\bar{\nabla}_{\mu}B\bar{\nabla}_{\nu}B \right)
\nonumber \\
+e^{-2B}f_{\mu\nu} \hat{Q} \\
R_{\mu b} =\frac{m}{2} ( D-2) \bar{\nabla}_{\mu} B
 \frac{\nabla_b{H}}{H} \\
R_{ab} =\hat{R}_{ab} - e^{2B} H^{n-m}\delta_{ab} 
\left( \bar{\nabla}^2 B + d_T \left( \bar{\nabla} B
\right)^2 \right) ,
\end{eqnarray}
where an overbar identifies terms that are calculated 
with the world--volume metric, $f_{\mu\nu}$. The 
quantity, $\hat{Q}$, represents a sum of terms 
depending on the function $H$ and its first and second derivatives. 
This sum is identical to the one that is obtained 
in the Ricci--flat limit, where $B=0$. 
Likewise, $\hat{R}_{ab}$ represents the 
transverse components of the 
Ricci tensor calculated from the  metric (\ref{brane1}).

We proceed to search for 
solutions to the field equations (\ref{field1})--(\ref{field3}) 
for the {\em ansatz} (\ref{branecos}). 
We  assume that the dilaton field 
has a separable form such that $\Phi (x, y) =\Phi_1 (x) +\Phi_2 (y)$, 
where the transverse--dependent part is given by 
the right--hand side of Eq. (\ref{brane2}). Moreover, 
we assume that the field strength satisfies
Eq. (\ref{brane3}). 
The introduction of a modulus field, $B$, 
leads to a non--trivial, off--diagonal component 
of the $D$--dimensional 
Ricci tensor. Nevertheless, the
$(\mu b)$--component of the 
Einstein equations (\ref{field1}) can be 
directly integrated to yield
the constraint
\begin{equation}
\label{sep}
\alpha \Phi_1=2(q-1) B
\end{equation}
relating the dilaton and modulus fields. 

The question that now arises is whether 
Eq. (\ref{sep})  
is compatible with the remaining field equations. 
We deduce by direct substitution 
that 
Eq. (\ref{field3}) is solved by Eqs. 
(\ref{brane2}), (\ref{brane3}) and (\ref{branecos}). 
Furthermore, by imposing the constraint 
\begin{equation}
\label{Bfield}
\bar{\nabla}^2 B + d_T \left( \bar{\nabla} B \right)^2 =0
\end{equation}
on the modulus field, we find that the 
$(ab)$--component of the Einstein 
equations (\ref{field1}) and the 
dilaton field equation (\ref{field2}) 
are also solved by Eqs. (\ref{brane2}), 
(\ref{brane3}) and (\ref{branecos}). 
Finally, the $(\mu \nu)$--component of Eq. (\ref{field1}) 
is solved by the same conditions if the 
Ricci tensor of the world--volume metric satisfies
\begin{equation}
\label{Efield}
\bar{R}_{\mu\nu} = d_T  \bar{\nabla}_{\mu\nu} B + 
\left( d_T +\frac{2(q-1)^2}{\alpha^2} \right) \bar{\nabla}_{\mu} B
\bar{\nabla}_{\nu} B   .
\end{equation}

Eqs. (\ref{Bfield}) 
and (\ref{Efield}) may be expressed in a more 
familiar form by performing the conformal 
transformation
\begin{equation}
\label{ct}
\tilde{f}_{\mu\nu} = \Omega^2 f_{\mu\nu} , \qquad 
\Omega^2 \equiv e^{2d_TB/(d_W-2)} 
\end{equation}
on the 
world--volume metric 
and rescaling the modulus field, $B\equiv Q^{-1} \chi$, where
\begin{equation}
\label{Bdefinition}
Q \equiv 
\sqrt{2} \left( q+1 +\frac{2(q-1)^2}{\alpha^2} 
+\frac{(q+1)^2}{d_W -2} \right)^{1/2}   .
\end{equation}
This implies that 
\begin{eqnarray}
\label{conformal1}
\tilde{R}_{\mu\nu} =\frac{1}{2} \tilde{\nabla}_{\mu}
\chi \tilde{\nabla}_{\nu} \chi \\
\label{conformal2}
\tilde{\nabla}^2 \chi =0
\end{eqnarray}
and Eqs. (\ref{conformal1}) and 
(\ref{conformal2}) represent the $d_W$--dimensional 
field equations for a massless scalar field minimally coupled 
to Einstein gravity. 

Thus, modulo a solution to Eqs. (\ref{conformal1}) and 
(\ref{conformal2}), we have found a class 
of solutions to the supergravity field equations 
(\ref{field1})--(\ref{field3}) that reduce to 
the Ricci--flat branes (\ref{brane1})--(\ref{brane3}) 
in the limit where the dilaton field 
is constant on the world--volume. Since the 
dependence of these solutions on the transverse coordinates 
is {\em identical} to that of the Ricci--flat limit, they 
may be interpreted as brane cosmologies, where the curvature 
of the brane is induced by the variation of the dilaton field
over the world--volume. 

In particular, 
the D$p$--brane cosmologies have a metric given in 
the Einstein frame by 
\begin{equation}
ds_{{\rm D}p}^2 =H^{(1-q)/8}f_{\mu\nu}dx^{\mu}dx^{\nu}
+H^{(9-q)/8} e^{\epsilon \Phi_1}\delta_{ij}dy^idy^j  ,
\end{equation}
where $\epsilon \equiv (5-q)/[2(q-1)]$, $\Phi_1 =
2B/\epsilon$, $e^{\Phi_2} = H^{(q-5)/4}$ and 
$\{ f_{\mu\nu}, B \}$ solve Eqs. (\ref{Bfield}) 
and (\ref{Efield}). The 
corresponding  metric in the string frame, 
$g^{(s)}_{AB} = \Theta^2
g_{AB}$, where $\Theta^2 \equiv e^{\Phi /2}$, is 
given by 
\begin{eqnarray} 
\label{RRstring}
ds^2_s =H^{-1/2} e^{\Phi_1 /2} f_{\mu\nu}dx^{\mu}dx^{\nu} 
\nonumber \\
+H^{1/2} e^{2\Phi_1/(q-1)} \sum_{j=1}^{d_T} 
dy_j^2  .
\end{eqnarray}

In effect, solutions of this type exist because 
Eqs. (\ref{field1})--(\ref{field3}) can 
each be separated into 
a sector  that depends only on the world--volume 
coordinates and a sector that 
depends only on the transverse coordinates. 
If the separation constants are then 
set to zero, the latter sector 
reduces to the field equations that arise 
in the Ricci--flat limit.

The dilaton and modulus fields must vary in direct proportion
to one other and the constant of proportionality
depends on the degree of the form field and its 
coupling to the dilaton. However, it is independent of the 
dimensionality of spacetime. 
There are two cases where the world--volume must remain Ricci--flat, 
however, at least within the context of the assumptions 
made above. The dilaton field must 
depend only on the radial 
coordinate of the transverse space in the 
case of a $(D-3)$--brane $(q=1)$ or when it 
is not directly coupled to the form field
$(\alpha =0)$.
 
In the following Section, 
we consider some of the above brane cosmologies from 
an eleven--dimensional, M--theoretic perspective. 

\section{Eleven--Dimensional Interpretations}

\setcounter{equation}{0}

\subsection{D8--Brane Cosmology}

An important brane 
that has received considerable attention 
is the D8--brane \cite{D8} 
of Romans' massive IIA theory \cite{Romans}. 
This domain wall is supported by a $0$--form coupled to the dilaton 
in Eq. (\ref{action}) by $\alpha =5/2$. 
The cosmological version of this brane is given by
\begin{equation}
\label{D8cosmology}
ds^2_{\rm D8} = H^{1/8} f_{\mu\nu}dx^{\mu}dx^{\nu} 
+H^{9/8}e^{-5\Phi_1 /2}dy^2  ,
\end{equation}
where $e^{\Phi_2} =H^{-5/4}$ and $\Phi_1 = -5B/4$.

The eleven--dimensional 
origin of the D8--brane is presently unclear, although 
Hull has shown that it 
can be obtained by reducing M--theory on a torus bundle over a circle 
in the limit where the bundle size vanishes \cite{hullmassive}.
We now derive a 
solution to eleven--dimensional supergravity
that can be related to a cosmological version of the 
D8--brane. 
Standard compactification of vacuum M--theory on a non--dynamical
two--torus leads 
to a nine--dimensional theory of the form 
(\ref{action}), 
where the field strength corresponds to that of a massless axion field, 
$F_A =\nabla_A \sigma$. The dilaton and axion parametrize 
the ${\rm SL}(2,R)/{\rm U}(1)$ coset. The existence of this 
non--compact global symmetry of the action implies 
that a generalized Scherk--Schwarz  
compactification on a circle may then be performed 
\cite{ScherkSchwarz}, where the axion 
field has a linear dependence on the circle's coordinate. 
This introduces a mass parameter (cosmological constant) 
in the eight--dimensional theory. After a suitable rescaling 
of the moduli fields, 
the reduced action takes the form of Eq. (\ref{action}), where 
$D=8$ and 
the coupling between the scalar field and $0$--form is given by $\alpha 
= \sqrt{19/3}$ \cite{LP}. 

Thus, the corresponding 
domain wall (6--brane) cosmology is of the form 
\begin{equation}    
\label{6branecosmology}
ds_8=H^{1/6} f_{\mu\nu}dx^{\mu}dx^{\nu} +e^{- \sqrt{19/3} \Phi_1}
H^{7/6} dy^2  ,
\end{equation}
where $H(y)= 1+m|y|$,  $\Phi_2 = - \sqrt{19/12} \ln H$ and 
$m$ is a constant representing the slope 
parameter of the nine--dimensional axion, $\sigma (x^{\mu},y) 
=\sigma (x^{\mu}) +my$. The dilaton 
field, $\Phi$, 
is a linear combination of the three moduli fields, 
$\vec{\varphi} 
=(\varphi_1, \varphi_2, \varphi_3)$,
originating from the diagonal components of the compactifying metric, 
i.e., 
$\Phi = \sqrt{3/19} \vec{b}_{123} . \vec{\varphi}$, 
where $\vec{b}_{123} = \left( -3/2 , \sqrt{7/4} , 
\sqrt{7/3} \right)$ \cite{LP}. 

Following the prescription of L\"u and Pope \cite{LP}, 
the solution may be oxidised back to eleven dimensions. We find 
that  
\begin{eqnarray}
\label{M9brane}
ds_{11}^2 = e^{\Phi_1 /3\alpha} f_{\mu\nu} dx^{\mu}dx^{\nu} 
+He^{-6 \Phi_1 /\alpha} dy^2 \nonumber \\
+ He^{-2 \Phi_1 /\alpha} 
\left( dz^2_2+dz^2_3  \right) \nonumber \\
+H^{-1}e^{2\Phi_1 /\alpha} \left( 
dz_1 +mz_2dz_3 \right)^2  .
\end{eqnarray}
The compactifying dimensions in Eq. (\ref{M9brane}) 
form a torus bundle \cite{torus}:
\begin{equation}
\label{torusbundle}
ds_B^2 = R^2 dz_2^2 +\frac{1}{{\rm Im} \tau}
\left| dz_1 +\tau dz_3 \right|^2   ,
\end{equation}
where the $T^2$ fibre is spanned by the periodic coordinates $\{ 
z_1 , z_3 \}$, $R=H^{1/2} e^{-\Phi_1 /\alpha}$ is the 
circumference of the circular base space and  $\tau \equiv mz_2 +i
He^{-2\Phi_1 /\alpha}$ is the complex structure. These 
degrees of freedom depend on both the 
world--volume and transverse coordinates.

We now 
compactify the eleven--dimensional 
metric (\ref{M9brane}) in the $z_1$ direction, producing 
a type IIA D6--brane cosmology supported by the magnetic 
charge, $m$, of the two--form, $F_2=mdz_2 \wedge 
dz_3$. Conformally 
transforming to the string frame 
and employing a 
standard
T--duality transformation \cite{standard} in the $z_3$ direction
leads to the corresponding D7 type IIB solution. 
Finally, applying the massive T--duality rules of Ref. \cite{D8} 
in the $z_2$ direction
produces a D8--brane cosmology given, in the Einstein frame, by 
\begin{eqnarray}
\label{MD8}
ds^2_{\rm D8} = H^{1/8} \left[ e^{\bar{\sigma}_1/30}f_{\mu\nu} 
dx^{\mu}dx^{\nu} \right. \nonumber \\
\left. 
+e^{-\bar{\sigma}_1/10} \left( dz^2_2 +dz^2_3 \right) \right]
+H^{9/8} 
e^{-5\bar{\sigma}_1 /2} dy^2  ,
\end{eqnarray}
where $\bar{\sigma}_1 = 5\Phi_1/(2\alpha )$
is the world--volume dependent part of the ten--dimensional 
dilaton field, $\Phi_1$ is 
given in Eq. (\ref{6branecosmology}) 
and $f_{\mu\nu}$ is the six--dimensional 
metric solving Eqs. (\ref{Bfield}) and (\ref{Efield}), 
where $B=-\sqrt{19/12} \Phi_1$. The solution 
(\ref{MD8}) has at least two 
abelian isometries on the world--volume 
and reduces to the Ricci--flat D8--brane 
in the limit where $\bar{\sigma}_1$ vanishes \cite{Brecher}. 

\subsection{NS5--Brane Cosmology}

Another important brane of the 
type IIA theory is the NS5--brane 
supported by the 
NS-NS three--form. This is 
coupled to the dilaton field 
such that $\alpha = -1$.
The corresponding NS5--brane cosmology is therefore
of the form 
\begin{eqnarray}
\label{NS5brane}
ds^2_{\rm NS5} = H^{-1/4} f_{\mu\nu} dx^{\mu}dx^{\nu} 
\nonumber \\
+H^{3/4} e^{-\Phi_1 /2}\left( dy_1^2 + \ldots + dy_4^2 \right)   ,
\end{eqnarray}
where $e^{\Phi_2} = H^{1/2}$ and $\Phi_1 = -4 B$.
It is well 
known that the type IIA theory may be derived by compactifying 
$D=11$ supergravity on a circle, 
where the radius of the circle is related to the 
string coupling by $r_{11} =g^{2/3}_s =e^{2 \Phi /3}$ 
\cite{Witten,Town}. 
Thus, the ten--dimensional 
brane cosmology (\ref{NS5brane}) may be oxidized 
to eleven dimensions to yield
\begin{eqnarray}
\label{NS5M5}
ds^2_{\rm M5} = H^{-1/3} \left( e^{-\Phi_1 /6} f_{\mu\nu} 
dx^{\mu} dx^{\nu} \right) \nonumber \\
+H^{2/3} \left( 
e^{-2 \Phi_1 /3} \delta_{ij} dy^idy^j +e^{4\Phi_1 /3} dz^2 \right)  ,
\end{eqnarray}
where $z$ is the coordinate of the eleventh dimension. 

Eq. (\ref{NS5M5}) represents a new 
solution to the $D=11$ supergravity equations of motion and 
may be interpreted as a M5--brane cosmology, 
where both the world--volume and transverse spaces are curved due  
to the dilaton's dependence on the world--volume coordinates. 
Indeed, the transverse space is no longer conformally 
flat in this case. Since the eleventh dimension 
becomes large in the strongly coupled limit, 
an equivalent interpretation of this 
solution is given in terms of 
a strongly--coupled NS5--brane cosmology
where the extra dimension is part of the transverse 
space. 

The NS5--solution  is also related to a 
D5--brane cosmology of the type IIB theory by 
S--duality \cite{HullIIB}. In type IIB supergravity, 
the dilaton and RR scalar field, $\lambda$,
parametrize the ${\rm SL}(2,R)/{\rm U}
(1)$ coset. Consequently, the theory exhibits a global 
${\rm SL}(2,R)$ symmetry \cite{SW}. The transformation is equivalent 
to the complex scalar field $\kappa \equiv \lambda +ie^{-\Phi}$ 
undergoing a fractional linear 
transformation: $\bar{\kappa} =(A\kappa +B)/
(C \kappa +D)$, where $AD-BC=1$. The Einstein--frame metric 
is a singlet under this transformation and the 
two--form potentials transform as a doublet. The
NS5 and D5 solutions are related by the special 
transformation  $A=D=0$ and $C=-B=1$ and this 
relates a strongly--coupled solution to a weakly coupled one since the 
sign of the dilaton field is reversed. 
It follows, therefore, that a more general type IIB brane cosmology 
may be generated from a seed NS5--brane 
solution by applying a global ${\rm SL}(2,R)$ symmetry 
transformation. This  produces a non--trivial 
scalar and 2--form potential in the RR sector. 
Furthermore, the 
dilaton field of the dual solution is given by 
$e^{\bar{\Phi}} = C^2e^{-\Phi} +D^2 
e^{\Phi}$ and cannot be separated into world--volume and 
transverse--dependent parts.

\section{Ho\v{r}ava--Witten Cosmology}

\setcounter{equation}{0}

Thus far, we have considered brane cosmologies within the context of the 
type II theories. However, the ${\rm E}_8 \times {\rm 
E}_8$ heterotic string theory has been favoured from a 
phenomenological perspective and it is therefore 
important to discuss its cosmological consequences. 
The strongly coupled limit of this theory 
is M--theory on an orbifold, $S^1/{\rm Z}_1$, 
and compactification 
on a Calabi--Yau three--fold leads to 
a gauged, five--dimensional supergravity theory 
with two four--dimensional boundaries. 
For the purposes of the present discussion, it is sufficient to 
consider a consistent truncation of this
theory that includes the 
breathing mode of the Calabi--Yau space, $\Phi$,
and a massless scalar field, $\sigma$, arising
from  the universal hypermultiplet. The action 
is given by
\begin{eqnarray}
\label{hwaction}
S=\int d^5 x \sqrt{|g|} \left[ R -\frac{1}{2} \left( 
\nabla \Phi \right)^2 -\frac{1}{2} e^{-\Phi} \left( \nabla 
\sigma \right)^2 \right. 
\nonumber \\
\left. 
-\Lambda e^{-2\Phi} \right] + 
\sum_{i=1}^2   (-1)^i\sqrt{24}\Lambda
\int d^4x \sqrt{\left| g_i \right|}
e^{-\Phi} .
\end{eqnarray}

The potential term  
in Eq. (\ref{hwaction}) is due to the non--trivial flux of the 
four--form field strength on four--cycles of the Calabi--Yau space
and it  
supports a solitonic 3--brane (domain wall) solution \cite{LOSW}.
This 3--brane has an eleven--dimensional interpretation 
in terms of 5--branes that are located on the 
ten--dimensional 
orbifold planes, where two of the dimensions 
are wrapped around 
a Calabi--Yau two--cycle. 

Cosmological brane solutions 
in Ho\v{r}ava--Witten theory 
have been found previously for a trivial axion field 
\cite{LOW,Reall,Lidsey,hwcos}. The five--dimensional metric is  given by 
\begin{equation}
\label{hwmetric}
ds^2_{\rm HW} = H f_{\mu\nu} dx^{\mu}dx^{\nu}  +H^4 e^{2B} dy^2 ,
\end{equation}
where $H=1+(2\Lambda /3 )^{1/2} |y|$, the breathing mode, 
$\Phi = \Phi_1(x) +\Phi_2 (y)$, 
is given by
\begin{equation}
\label{hwdilaton}
\Phi_1 = B , \qquad \Phi_2 =3 \ln H
\end{equation}
and the world--volume metric, $f_{\mu\nu}$, is determined by the 
effective field equations
\begin{eqnarray}
\label{hwfield1}
\bar{R}_{\mu\nu}  = \bar{\nabla}_{\mu\nu} B +
\frac{3}{2} \bar{\nabla}_{\mu} 
B \bar{\nabla}_{\nu} B \\
\label{hwfield2}
\bar{\nabla}^2 B +\left( \bar{\nabla} B\right)^2  =0 .
\end{eqnarray}

We 
now consider the effects of introducing 
the axion field, $\sigma$. We assume 
that
the metric and breathing mode are given by Eqs. 
(\ref{hwmetric}) and (\ref{hwdilaton}), respectively. 
The $(\mu y)$--components of the Einstein 
field equations are still solved by the separable ansatz 
(\ref{hwdilaton})  if the 
axion field is independent of the world--volume coordinates. 
We 
therefore assume that it depends only on
the transverse dimension, $y$. Its field equation 
then admits the first integral 
\begin{equation}
\label{firstintegral}
\sigma' =AH^3  ,
\end{equation}
where 
a prime denotes 
differentiation 
with respect to $y$ and 
$A$ is an arbitrary constant of integration. 
The $(\mu \nu )$--components of the 
Einstein field equations are
solved 
as before provided that the 
Ricci tensor of the world--volume satisfies 
Eq. (\ref{hwfield1}). However,  the 
equation of motion for the 
breathing mode acquires an additional 
term due to the axion field. It is 
solved if 
\begin{equation}
\label{sfield1}
\bar{\nabla}^2 B + \left( \bar{\nabla} B \right)^2 =-
\frac{A^2}{2} e^{-3B}
\end{equation}
and it can be shown that the
$(yy)$--component of the Einstein equations is also 
solved if Eq. (\ref{sfield1}) is satisfied. 

Hence, the compactified heterotic M--theory action (\ref{hwaction}) 
admits a curved 
domain wall cosmology of the form given by
Eq. (\ref{hwmetric}), where the axion field 
satisfies Eq. (\ref{firstintegral}). 
The cosmological 
expansion of the brane is determined by the conditions 
(\ref{hwfield1}) and 
(\ref{sfield1}). 
Performing the conformal 
transformation 
\begin{equation}
\label{hwconformal}
\tilde{f}_{\mu\nu} = \Theta^2 f_{\mu\nu} , \qquad 
\Theta^2 \equiv e^{B}
\end{equation}
and field redefinition $B=\chi /2$
implies that these conditions are equivalent 
to 
\begin{eqnarray}
\label{hwequiv1}
\tilde{R}_{\mu\nu} =\frac{1}{2} \tilde{\nabla}_{\mu} \chi
\tilde{\nabla}_{\nu} \chi +\frac{A^2}{4} \tilde{f}_{\mu\nu}
e^{-2\chi} \\
\label{hwequiv2}
\tilde{\nabla}^2 \chi =-A^2 e^{-2\chi} .
\end{eqnarray}
Eqs. (\ref{hwequiv1}) and (\ref{hwequiv2}) 
may be 
interpreted as the four--dimensional 
field equations for a minimally coupled scalar field, $\chi$,
that self--interacts through an exponential potential
$V =(A^2/2)e^{-Q\chi}$, where $Q=2$. The momentum of the axion field 
in the orbifold direction manifests itself 
to an observer on the brane as  a self--interaction 
potential for the breathing mode of the 
Calabi--Yau space \cite{Reall}. For an exponential 
potential of this type, the 
late--time 
attractor for the spatially flat 
Friedmann--Robertson--Walker (FRW) cosmology
is a power law, $a \propto t^{1/Q^2}$, for 
$Q^2 \le 3$, otherwise it is $a \propto t^{1/3}$ \cite{liddle}. 
In this Ho\v{r}ava--Witten model, 
$Q^2 = 4$, and the latter situation therefore arises. 
Hence, the unique late--time attractor 
is non--inflationary, although it is interesting that 
a potential for the 
breathing mode can be generated in this fashion. 

\section{Discussion and Conclusion}

\setcounter{equation}{0}

The solitonic D$p$-- and 
M--branes of string and M--theory have played a central role in 
establishing the duality relationships that exist 
between the different theories. 
A necessary condition for 
a brane to be interpreted cosmologically is that its 
world--volume  should be non--static and 
curved due to the existence 
of matter fields varying dynamically on the brane. 
We 
have found that at the level of the 
supergravity field equations, the world--volume 
of many of these branes becomes curved 
when the dilaton has 
a non--trivial dependence on the world--volume coordinates
and is related to the transverse dimensions in an 
appropriate way. 
In particular, we have presented
a  cosmological version of the M5--brane, where 
both the world--volume and
transverse spaces are curved. 
This solution 
represents the strongly--coupled limit of an NS5--brane cosmology. 
An eleven--dimensional interpretation was also given for 
a cosmological D8--brane of the massive type  IIA theory. 
Finally, a class of strongly--coupled braneworlds was found 
in heterotic M--theory compactified on 
a Calabi--Yau space. 

Moreover, the geometry of the brane 
world--volume was kept 
arbitrary in the analysis and was not 
restricted to the spatially isotropic FRW metrics.
This is important since the 
effects of spatial anisotropy and 
inhomogeneity may have been significant in the 
very early universe.  
The problem of solving the field equations 
(\ref{field1})--(\ref{field3}) was reduced to solving  
Einstein gravity minimally coupled to a massless 
scalar field and this system has been extensively 
studied in the literature.

We conclude by considering the possibility that 
the kinetic 
energy of the dilaton 
field can drive an epoch of inflationary 
expansion 
on the D$p$--branes. In the standard, pre--big 
bang scenario, the simplest 
solution is that of 
the spatially flat, homogeneous Bianchi I model
defined over $t<0$. (For a review, see, e.g., Ref. \cite{lwc}). This 
is the time--reversal 
of the `rolling radii' solution of  Mueller \cite{mref}
and represents 
a generalization of the Kasner solution \cite{kasner}. The string frame metric 
and dilaton field are given by
\begin{equation}
\label{m1}
ds^2=-dt^2 +\sum_{i=1}^9 (-t)^{2\beta_i} dz_idz^i
\end{equation}
and 
\begin{equation}
\label{m2}
\Phi=-\frac{1}{2} \left( 1-\sum_{i=1}^9 \beta_i \right) \ln  (-t)  ,
\end{equation}
respectively, where the constants, $\{ \beta_i \}$, satisfy the constraint
equation
\begin{equation}
\sum_{i=1}^9 \beta_i^2 =1  .
\end{equation}

Given the nature of Eq. (\ref{RRstring}), 
we consider D$p$--brane cosmologies 
of the form 
\begin{eqnarray}
\label{meullersol}
ds^2 = H^{-1/2} \left( -dt^2 +\sum_{i=1}^{d_W-1} 
(-t)^{2\beta_i} dx^2_i \right) \nonumber \\
+H^{1/2} (-t)^{2\gamma} 
\sum_{j=1}^{d_T} dy_j^2  ,
\end{eqnarray}
where $\Phi_1$ depends only 
on cosmic time. 
The constraints on the exponents $\{ \beta_i , \gamma \}$ may 
be deduced by noting that the $q$--form 
field strength supporting the 
brane is non--trivial only in the 
transverse--dependent sector of the field equations 
(\ref{field1})--(\ref{field3}). Thus, the time--dependence 
of the metric and dilaton is given by the 
rolling radii solution (\ref{m1}) and (\ref{m2}): 
\begin{eqnarray}
\label{muellerphi}
\Phi_1 =-\frac{1}{2} \left( 1 -\sum_{i=1}^{d_W-1}
\beta_i -d_T \gamma
\right) \ln (-t) \\
\label{kasner}
\sum_{i=1}^{d_W-1} \beta_i^2 +d_T \gamma^2 =1  .
\end{eqnarray}

However, there is an additional constraint
on the exponents 
because the dilaton field is directly 
related to the transverse dimensions 
by the separability condition (\ref{sep}).
Comparison 
of  Eqs. (\ref{RRstring}) and (\ref{meullersol}) 
implies that 
\begin{equation}
\label{gammaconstraint}
1-\sum_{i=1}^{d_W-1} \beta_i =(3-q) \gamma
\end{equation}
and thus, for $q\ne 3$, 
\begin{equation}
\label{extraconstraint}
\sum_{i=1}^{d_W-1} \beta_i^2 + \frac{q+1}{(q-3)^2} \left( 1-
\sum_{i=1}^{d_W-1} \beta_i \right)^2 =1  .
\end{equation}

During dilaton--driven 
inflation, the string coupling increases 
in value. Eqs. (\ref{muellerphi}) and (\ref{gammaconstraint})
imply that a necessary condition 
for inflation is  $\gamma (q-1)<0$  
and there is a wide region of parameter space 
where inflation of this type can proceed 
on the brane. For 
example, let us consider the D8--brane cosmology (\ref{D8cosmology})
of the massive type IIA theory and, 
for simplicity, assume that five of the world--volume 
dimensions are independent of time and that the remaining 
three are isotropic, $\beta_i =\beta$. 
Eq. (\ref{extraconstraint}) then implies that 
$\beta =(1\pm \sqrt{33})/12$ and $\gamma = 
(3\mp \sqrt{33})/12$ and the negative root 
therefore 
leads to accelerated expansion as $t \rightarrow 0^-$. 

Finally, we remark that the 
Ricci--flat branes (\ref{brane1})--(\ref{brane3})  
are also directly relevant to cosmology 
as a consequence of a powerful embedding theorem due to 
Campbell \cite{campbell}. This theorem 
states that 
{\em any} analytic Riemannian space of dimension $n$ and 
signature $(1,n-1)$ can be locally and isometrically 
embedded in a Ricci--flat, Riemannian space of dimension $n+1$ 
and signature $(1,n)$. The embedding 
is established by solving a set of 
constraint equations that are compatible with 
the Gauss--Codazzi equations \cite{rtz}.
In particular, 
perfect fluid FRW cosmologies can be embedded in this fashion 
\cite{pf}. 
>From a five--dimensional point of view, 
the solution is interpreted as a shock wave 
travelling through time and the fifth dimension
\cite{wls}
and the non--trivial energy--momentum tensor 
is 
induced on the four--dimensional hypersurface
by relaxing the cylinder condition of 
Kaluza--Klein theory \cite{OW}. 
Since Campbell's theorem is independent 
of the dimensionality of the space, the 
procedure may be repeated an arbitrary number of times 
to embedded four--dimensional cosmologies 
in Ricci--flat branes of higher dimensions, such as the M5--brane. 
It would be interesting to explore such embeddings further. 

\acknowledgments
The author is supported by the Royal Society. 
We thank J. Gauntlett for helpful comments.

\end{document}